\documentclass{ws-procs9x6}

\begin{document}

\title{Quantum control, quantum information processing, and quantum-limited metrology with trapped ions}

\author{D. J. Wineland, D. Leibfried, M. D. Barrett\footnote{Present
address: \uppercase{P}hysics \uppercase{D}epartment,
\uppercase{U}niversity of \uppercase{O}tago, \uppercase{N}ew
\uppercase{Z}ealand} , A. Ben-Kish\footnote{Present address:
\uppercase{T}echnion, \uppercase{H}aifa, \uppercase{I}srael},\\ J.
C. Bergquist, R. B. Blakestad, J. J. Bollinger, J. Britton,\\ J.
Chiaverini, B. DeMarco\footnote{Present address: \uppercase{P}hysics
\uppercase{D}ept., \uppercase{U}niv. of \uppercase{I}llinois}, D.
Hume, W. M. Itano, M. Jensen,\\ J. D. Jost, E. Knill, J. Koelemeij,
C. Langer, W. Oskay, R. Ozeri,\\ R. Reichle, T. Rosenband, T.
Schaetz\footnote{present address: \uppercase{M}ax \uppercase{P}lanck
\uppercase{I}nst.\ for \uppercase{Q}uantum \uppercase{O}ptics,
\uppercase{G}arching, \uppercase{G}ermany}, P. O.
Schmidt\footnote{present address: \uppercase{I}nst.\ for
\uppercase{E}xperimental \uppercase{P}hysics, \uppercase{U}niversity
of \uppercase{I}nnsbruck, \uppercase{A}ustria},\\ and S. Seidelin}

\address{National Institute of Standards and Technology, Boulder CO 80305-3328\\
E-mail: djw@boulder.nist.gov}

\maketitle

\abstracts{We briefly discuss recent experiments on quantum
information processing using trapped ions at NIST.  A central theme
of this work has been to increase our capabilities in terms of
quantum computing protocols, but we have also applied the same
concepts to improved metrology, particularly in the area of
frequency standards and atomic clocks. Such work may eventually shed
light on more fundamental issues, such as the quantum measurement
problem.}

\section{Introduction}
In 1995, Ignacio Cirac and Peter Zoller described how an ensemble of
trapped ions could be used to implement quantum information
processing (QIP).\cite{CandZ}  Several experimental groups
throughout the world have pursued this basic idea, and although a
useful device still does not exist, ion-trappers are optimistic that
one can eventually be built. In part, this is because the ion-trap
scheme can satisfy the basic requirements for a quantum computer as
outlined by DiVincenzo\cite{divincenzo01}: (1) a scalable system of
well defined qubits, (2) a method to reliably initialize the quantum
system, (3) long coherence times, (4) existence of universal gates,
and (5) an efficient measurement scheme. Most of these requirements
have been demonstrated, and straightforward, albeit technically
difficult, paths to solving the remaining problems exist. In this
paper, we summarize recent trapped-ion QIP experiments carried out
at NIST, but note that similar work is currently being pursued at
Aarhus, Barcelona, Garching (MPQ), Innsbruck, LANL, London
(Imperial), Ontario (McMaster), Michigan, MIT, Oxford, Siegen,
Sussex, Teddington (NPL), and Ulm.

We describe how the system might be scaled up by use of an array of
interconnected trap zones and cite experimental implementation of
algorithms that utilize the basic elements of this scheme.  We then
summarize efforts devoted to construction of traps by use of methods
that are suitable for large-scale fabrication.  We briefly discuss
how QIP methods might be used in metrology, and finally suggest how
QIP studies might eventually shed light on fundamental issues of
decoherence.

\section{QIP with multiplexed ion trap arrays}
Although large numbers of ions can be cooled into regular arrays in
single traps, many of the practical $N$-ion gates ($N \geq 2$), such
as the original Cirac/Zoller two-ion gate,\cite{CandZ} require
addressing of individual ions and single modes (or a very small
number of modes) of ion motion.  Individual ion addressing can be
accomplished with focused laser beams as long as the ions aren't too
close together (or equivalently, as long as the mode frequencies are
not too high).  Mode addressing is usually accomplished by
spectrally isolating the mode(s) of interest (out of $3N$ possible
modes). This has the consequence that when the number of trapped
ions becomes large, the mode spectrum becomes so dense that spectral
isolation becomes impractical. Although the group at Innsbruck has
successfully implemented a number of interesting algorithms on
multiple ions in single trap zones by using focused laser beams for
individual qubit addressing (see their contribution to these
proceedings), as the number of ions increases further, and increased
gate speeds (proportional to mode frequencies) become more
important, such addressing will become more difficult.

Therefore, many groups are considering a multiplexed system of
trapping zones where only a small number of ions are confined in the
zones that are used for implementing gates. The sharing of quantum
information between zones might be accomplished by moving ion qubits
between zones,\cite{bible,kielpinski02} by moving an
information-carrying ``head" ion between zones,\cite{cirac00} by
coupling separated ions with photons as an
intermediary,\cite{devoe98} or by probabilistically creating
entangled pairs of separated ions via light coupling, which then act
as a computational resource to be used later.\cite{duan04}

\subsection{QIP in a linear ion trap array}
As a first step towards multiplexing, we have used a six-zone linear
array that is an extension of the three-zone trap reported
earlier.\cite{rowe02} Recent experiments with this device have
included demonstrations of quantum teleportation,\cite{barrett04}
quantum error correction,\cite{chiaverini04} quantum-dense
coding,\cite{schaetz05} and the quantum Fourier
transform.\cite{chiaverini05}

\begin{figure}
\centerline{\epsfxsize=4.1in\epsfbox{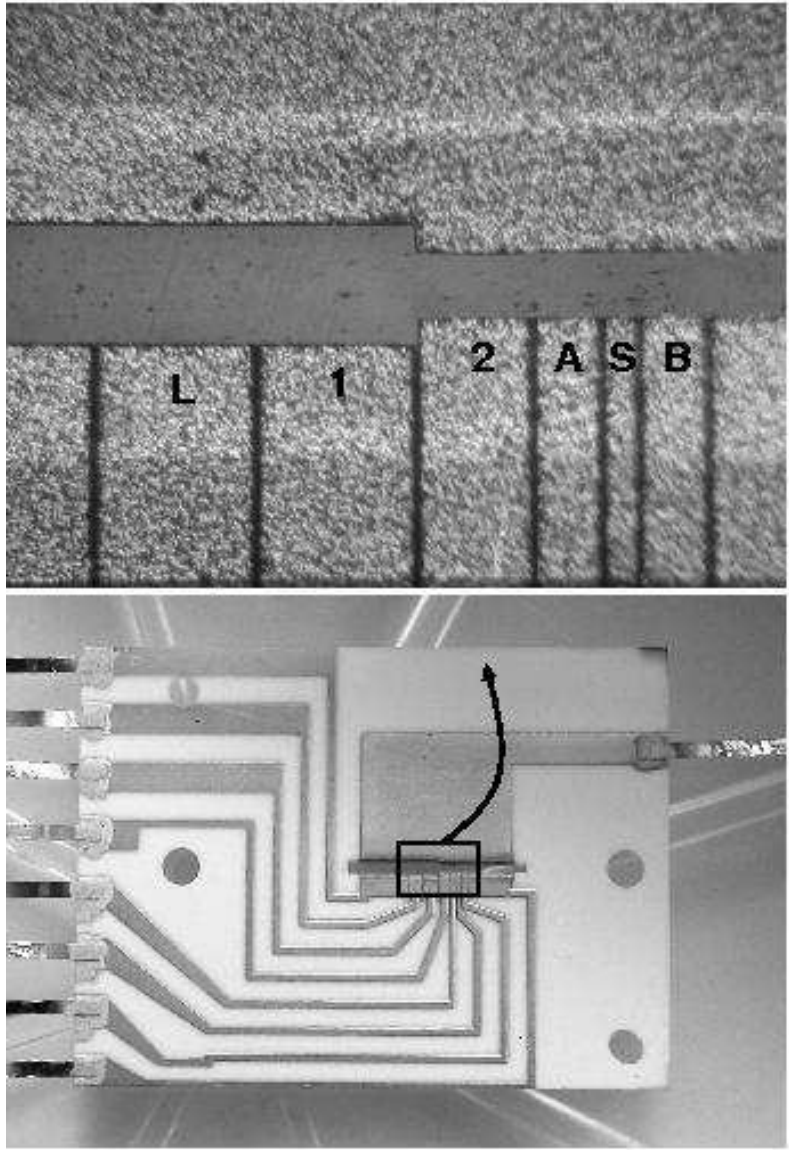}}
\caption{Photograph of one wafer of a six-zone linear trap array.
Two of these wafers, properly spaced, comprise the trap as described
in [8]. The lower part of the figure shows gold traces
(approximately 3 $\mu$m thick) deposited onto an alumina substrate
(lighter color). The upper figure is an expanded view of the boxed
section shown below. For the wafer shown, an RF potential ($\sim
200$ V at $\sim$ 150 MHz) is applied to the upper (continuous)
electrode. ``Control" potentials are applied to the eight segmented
electrodes. Varying the potentials on these electrodes in a
coordinated way enables ions to be moved between the six zones
located above the electrodes labeled L,1,2,A,S, and B. Zone L is the
``Loading" zone, whose width is relatively large to increase the
capture volume for beryllium atoms (emitted from a thermal source)
that are ionized (by electron impact) in this area. In most of the
algorithms demonstrated using this trap, zones A and B (``Alice" and
``Bob") were used to manipulate the internal states of qubits (with
laser beams overlapping those zones).  (Traps constructed by M. D.
Barrett and J. D. Jost) \label{6-ZONE}}
\end{figure}

These experiments required that entanglement between ions was
preserved when the ions were located in different zones. Referring
to Fig. \ref{6-ZONE}, entanglement was created in zone A, and the
ions were sent to zone S for separation.  Electrode S is relatively
narrow to facilitate separation of a single group of ions into
subgroups by inserting a potential wedge between selected ions. For
example, in the teleportation experiment on $^9$Be$^+$
ions,\cite{barrett04} three ions could be separated into a group of
two which were delivered to zone A, with the third ion delivered to
zone B. We optimized the separation to minimize the heating of the
ions delivered to zone A.  With a separation time of 200 $\mu$s, the
ions could be separated without error.  The axial center-of-mass
motion of ions in zone A (frequency $\sim$ 3 MHz) experienced a
kinetic energy increase corresponding to about 1 quantum, the
stretch mode had gained negligible kinetic energy, and the axial
motion in zone B gained about 10 quanta.  In the future, traps with
much smaller internal dimensions should enable shorter separation
times with negligible heating, due to the higher motional
frequencies and sharper separation potential wedge features.
However, with all other parameters held constant, smaller dimensions
will aggravate ion heating\cite{turchette00a} and sympathetic
cooling will likely be required to maximize gate
fidelity.\cite{bible,kielpinski02}

Other recent experiments in these traps (that did not require
multiple zones) included investigations of spontaneous emission
decoherence during Raman transitions\cite{ozeri05} and a long-lived
($\tau_1$, $\tau_2 >$ 10 s) qubit memory based on first-order
magnetic field-insensitive transitions.\cite{langer05}

\subsection{Future ion trap arrays}
For manipulating very large numbers of ions with high gate speeds,
it appears that new types of trap construction methods, including
two-dimensional layouts, will be required.  Since (two-qubit) gate
speed is proportional to the ions' motional frequencies, which are
in turn proportional to (electrode dimensions)$^{-2}$, we would like
to implement traps with dimensions smaller than those of the traps
indicated in Fig. \ref{6-ZONE}. Such gold coated alumina electrode
structures,\cite{rowe02,barrett04,deslauriers04} have a size
limitation from the fact that the laser-machined cuts in the wafers
are limited to a width of around 20 $\mu$m.

To overcome this limitation, it should be possible to take advantage
of MEMS fabrication techniques, where significantly smaller
structures can be fabricated. If this is done, we must of course
worry about ion motional heating, which increases as the electrodes
become smaller.\cite{turchette00a}  An obvious construction material
would be silicon; however, with typically available substrates, RF
loss at the trap drive frequency appears to be a problem.

\begin{figure}[ht]
\centerline{\epsfxsize=4.1in\epsfbox{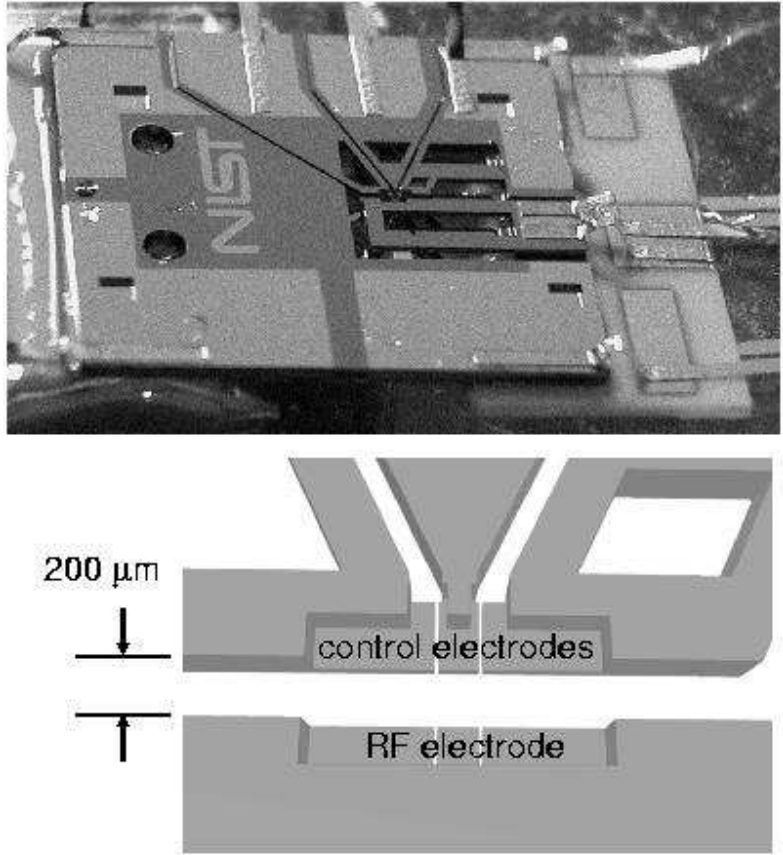}}
\caption{The photograph shows a single-zone two-layer trap of the
type described in Ref. [8]  The bottom part of the figure shows a
schematic of the trapping region for one of the trap electrode
wafers, which are fabricated from boron-doped silicon. Laser cooled
$^{24}$Mg$^+$ ions have been confined in this trap (constructed by
J. Britton, NIST).\label{Joes-trap}}
\end{figure}

At NIST we constructed a single-zone two-layer trap of the type
described in Ref. [8] whose electrodes were made of commercially
available boron-doped silicon (Fig. \ref{Joes-trap}). In this
apparatus, we trapped and laser cooled $^{24}$Mg$^+$ ions. Electrode
features as small as 5 $\mu$m were defined by use of
photolithography and industry standard silicon deep reactive ion
etching (DRIE Bosch process). Structural support and spacing of the
electrodes was provided by a borosilicate glass thermally matched to
silicon and attached to the electrodes by anodic
bonding.\cite{kielpinski_thesis}  Such an approach is applicable to
the fabrication of many-zone large-scale traps including planar
traps (below) since the number of processing steps does not increase
with the number of zones in the array.  In a different approach, the
University of Michigan group has built a two-layer trap with GaAs
electrodes and AlGaAs insulators\cite{madsen04} and observed
trapping of Cd$^+$ ions.\footnote{C. Monroe, Univ. of Michigan,
private communication} A three-layer geometry\cite{kielpinski02} has
been implemented for Cd$^+$ ions\cite{deslauriers04} and geometries
that would optimize the separation of ions into separate groups have
been studied.\cite{home04} Sandia researchers have fabricated arrays
of very small ($\sim 1\ \mu$m) three-dimensional trap
structures,\cite{blain04} which also might be configured for QIP.

Borrowing from the groups pursuing magnetic waveguide traps for
neutral atoms, linear traps based on electrodes confined to a
surface might also be considered.\cite{chiaverini05b} Such ``planar"
traps would be relatively easy to fabricate on a large scale and
would permit on-board electronics beneath the electrode
surface.\footnote{R. Slusher, Lucent, private communication}

In addition to finding a way to construct large-scale trap arrays, a
way to multiplex laser beams must be sought.  It might be possible
to use miniature steerable mirrors based on MEMS technology for this
purpose.\footnote{O. Blum Spahn, Sandia National Labs, private
communication.}$^,$\footnote{J. Kim, Duke University, private
communication.}  Miniature, large-solid-angle photon detectors
(possibly without optics) located very near trapping zones may be
essential for highly parallel detection as required in error
correction.

\section{QIP applied to metrology}
In the Time and Frequency Division of NIST, we have been interested
in applying the methods of QIP to metrology, in particular, to
improve the signal-to-noise ratio in spectroscopy and atomic clocks.
For this purpose, we take advantage of entanglement.   The
improvement obtained from ``spin-squeezed"
states,\cite{wineland92,kitagawa93,wineland94} where the operator of
the effective mean spin vector is measured, has been demonstrated
for two ions.\cite{meyer01}  We have also demonstrated the gain in
signal-to-noise ratio with certain states in combination with other
operators such as the variance and parity.\cite{bollinger96,meyer01}
More recently, we have extended a two-ion phase gate\cite{didi_gate}
to implement a form of Ramsey spectroscopy where each of the two
conventional Ramsey $\pi$/2 pulses are replaced with a rotation and
one-step phase gate.\cite{leibfried04} Starting with all ions in the
state $| \downarrow \rangle$, the first modified ``$\pi/2$" pulse
generates a generalized GHZ state\cite{GHZ} or
``Schr{\"o}dinger-cat" state of the form
$\textstyle{\frac{1}{\sqrt{2}}}[|\!\!\downarrow\rangle_{1}|\!\!\downarrow\rangle_{2}
\cdot\cdot\cdot |\!\!\downarrow\rangle_{N} + e^{i\beta}
|\!\!\uparrow\rangle_{1}|\!\!\uparrow\rangle_{2} \cdot\cdot\cdot
|\!\!\uparrow\rangle_{N}]$.  During the Ramsey free-precession
interval $T$, the relative phase of the two components of the
wavefunction $\beta = N(\omega_0 - \omega)T$, where $\omega$ is the
frequency of the probe oscillator and $\omega_0$ is the resonance
transition frequency, advances $N$ times faster than that of a
single atom. This is the main reason for the increase in
spectroscopic resolution. After application of the two modified
Ramsey pulses, the equivalent net spin vector is measured in the
$|\!\downarrow \rangle$, $|\!\uparrow \rangle$ basis. In ideal
circumstances, all ions are measured to be in either all $| \uparrow
\rangle$ states with probability $P_{\uparrow} =
\textstyle{\frac{1}{2}}[1 + \cos{N(\omega_0 - \omega)T}]$ or all $|
\downarrow \rangle$ states with probability $P_{\downarrow} = 1 -
P_{\uparrow}$, cases that are relatively easy to distinguish.
Although the fringes occur $N$ times faster, the gain in
signal-to-noise ratio is limited to $\sqrt{N}$ compared to the case
of $N$ unentangled particles, because the $N$ unentangled particles
yield a signal from $N$ separate systems, where the ``projection"
noise\cite{itano93} averages down as $N^{-1/2}$. Although the
experimentally observed gain was limited to less than $\sqrt{N}$, we
were able to demonstrate a signal-to-noise ratio better than could
be obtained in a perfect experiment on unentangled ions, first on
three ions\cite{leibfried04} and more recently on up to six
entangled ions.\cite{leibfried_6ions}

QIP might also be used to improve detection.  In one application
relevant for frequency standards, it was shown that transitions in a
``clock" ion can be detected in a simultaneously-trapped ``logic"
ion by mapping the internal state of the clock ion onto the logic
ion (with elementary quantum logic operations) where it is easily
detected.\cite{schmidt05}  In a more general context, detection
sensitivity of quantum systems can be improved in certain situations
by use of elementary quantum logic operations on the system to be
measured, in conjunction with ancilla particles that are also
measured.\cite{schaetz05}

\section{QIP and the ``measurement problem"}
By the measurement problem, we mean the difficulty that arises
because we live in a world that predicts definite outcomes (e.g.,
bits in our PCs are either 0 or 1) whereas quantum mechanics alone,
in general leaves the world in superposition states.  In addition to
the simple collapse postulate, many attempts have been made to
resolve the problem with ideas that include concepts such as ``many
worlds," decoherence theory, an as-of-yet unseen collapse mechanism,
or simply that the theory of quantum mechanics is only a
computational tool that allows prediction of classical outcomes (for
a recent review, see for example the paper by
Leggett\cite{leggett02}).

Given the unresolved state of affairs on the measurement problem, it
seems interesting to press the issue experimentally {\em -} that is,
can we realize larger and larger entangled superposition states that
begin to approach our more macroscopic world where such states
aren't observed?  The paper by Leggett suggests one measure for
approaching the classical world in which the number of elementary
particles involved in a superposition state is of primary
importance.\cite{leggett02} At this stage, since we really don't
know what the important parameters are, we might cook up alternative
measures that play more to the strengths of atomic physics and
quantum optics. With atomic ions, we can emphasize the aspects of
entanglement and duration. For example, we might take as a figure of
merit the product of the number of particles in a GHZ state (since
its phase sensitivity is $N$-fold larger than that of a single
particle) times the duration of the state.  A start in this
direction is that a six particle approximation to a GHZ state was
observed to last longer than approximately 50
$\mu$s.\cite{leibfried_6ions} Note also that superpositions of the
(phase-insensitive) Bell states $\Psi_{\pm} =
{\textstyle{\frac{1}{\sqrt{2}}}}(| \downarrow \rangle_1| \uparrow
\rangle_2 \pm | \uparrow \rangle_1| \downarrow \rangle_2)$ have been
observed to last for durations exceeding 5 s in $^9$Be$^+$
ions\cite{langer05} and even longer for Ca$^+$ ions (see the paper
by the Innsbruck group in these proceedings). Whatever your favorite
measure is, it seems likely that as the quest to make a large-scale
QIP machine progresses, states that look more and more like
Schr{\"o}dinger's cat will be produced {\em -} or not, if some
fundamental source of decoherence is discovered!

\section*{Acknowledgments} We thank E. Donley and S. Jefferts for comments on the manuscript.
This work was supported by the US
National Security Agency (NSA) and the Advanced Research and
Development Activity (ARDA) under contract number MOD-7171.05. The
work was also supported, in part, by the US Office of Naval Research
(ONR) and NIST.  This manuscript is a publication of NIST and is not
subject to U. S. copyright.

\end{document}